\documentclass{PoS}
\usepackage{etex}
\usepackage[export]{adjustbox}
\usepackage{mcite}
\usepackage{epsfig}

\usepackage{url}

\usepackage{color}

\usepackage[utf8]{inputenc}
\usepackage[T1]{fontenc}
\usepackage{lmodern}
\usepackage{amsfonts}
\usepackage{supertabular}
\usepackage{color}
\usepackage{textcomp}
\usepackage{amsmath}
\usepackage{amssymb}
\usepackage{graphicx}
\usepackage{lineno}
\usepackage{xcolor}
\usepackage{wrapfig}
\usepackage{subfigure}
\usepackage{type1cm}
\usepackage{ulem}
\usepackage{amsmath}
\usepackage{amssymb}
\usepackage{amscd}
\usepackage{xspace}
\usepackage{upgreek}
\usepackage{color}
\usepackage{calc}
\usepackage{ifthen}
\usepackage{cite}
\usepackage{enumitem} 
\usepackage{multirow}
\usepackage{amssymb}
\usepackage{mathtools}
\usepackage{graphicx}
\usepackage{ifthen}
\usepackage{caption}
\usepackage{cleveref}
\usepackage{floatrow}

\newcommand{\eVdist}{\kern-0.06667em}
\newcommand{\GeV}{{\text{Ge}\eVdist\text{V\/}}}

\newcommand{\fbi}{\,\text{fb}^{-1}}

\newcommand{\HERAII}{{ HERA-II}\xspace}
\newcommand{\HERAI}{{ HERA-I}\xspace}
\newcommand{\HERA}{{ HERA}\xspace}

\newcommand{\DESY}{{DESY}\xspace}

\newcommand{\ZEUS}{{ ZEUS}\xspace}

\newcommand{\ZEVIS}{{ZeVis}\xspace}

\newcommand{\MC}{{MC}\xspace}

\newcommand{\CN}{{CN}\xspace}

\newcommand{\INSPIRE}{{InSpire}\xspace}
\mathchardef\mhyphen="2D

\title{The ZEUS long term data preservation project}

\ShortTitle{The ZEUS long term data preservation project}

\author{\speaker{Andrii Verbytskyi}\thanks{On behalf of the ZEUS 
collaboration}\\
        Max-Planck Institut f\"{u}r Physik (Werner Heisenberg Institut),
        F\"{o}ringer Ring 6,  M\"{u}nchen 80605, Germany\\
        E-mail: \email{andrii.verbytskyi@mpp.mpg.de}}

 \abstract{The ZEUS data preservation (ZEUS DP) project assures continued
 access to the data and documentation related to the experiment. It aims
  to provide the ability to continue the generation of valuable 
 scientific results from these data in the future. This talk presents 
 analysis opportunities within the ZEUS DP project and its benefits for 
 the physics community. The implications of the preserved data are 
 discussed in the context of current data analyses, verification of 
 contemporary and future theoretical models and the planning of future 
 experiments and analyses. The detailed documentation on the data 
 preservation effort can serve as an example for the data preservation 
 efforts in current and future experiments.}

 \FullConference{XXIV International Workshop on Deep-Inelastic Scattering
 and Related Subjects\\ 11-15 April, 2016\\ DESY Hamburg, Germany}

\begin{document}
 \section{Introduction}
 \label{sec:int}
 The data collected by high energy physics experiments are crucial for 
 the understanding of particle physics. Among the many high energy 
 physics projects the experiments on the  \HERA collider hold a special 
 place, since it was (and still is) the world's only $ep$ collider. 
 Therefore the \HERA data are unique and considered for the preservation.

 The data preservation effort aims to ensure the long-term availability 
 of the data after the end of the data taking and funding periods of the 
 experimental collaborations. In this way it can increase the physics 
 potential of the experiments, allowing long-term data analysis, re-using 
 and re-analysing the data, combining results between experiments and 
 using the data for education, training and outreach purposes. 
 The data preservation efforts for the HERA experiments are part of a 
 larger data preservation effort~\cite{Akopov:2012bm}.
 \section{HERA and ZEUS}
 \label{sec:HERA}
 The Hadron Electron Ring Anlage (\HERA) was a  ring accelerator at DESY 
 in Hamburg~\cite{zeus:1993:bluebook}, so far the only 
 electron(positron)-proton collider in the world.
 It was running between 1991 and 2007 with a break in 2001-2002.
 During the   break the 
 \HERA ring was upgraded: the luminosity was increased  and electron 
 polarisation systems were installed. The data 
 taking period before the break is referred to as \HERAI, while the one 
 after the break is referred to as \HERAII. During the all time of running 
 the electron(positron) beam energy was close to $27.5\GeV$, while the 
 proton beam energies varied, being $820\GeV$ and $920\GeV$ for \HERAI, 
 $920\GeV$,  $575\GeV$ and $460\GeV$ for \HERAII.
 \begin{figure}[phtb]\vfill
 \begin{center}
 \adjincludegraphics[width=0.4\linewidth]
 {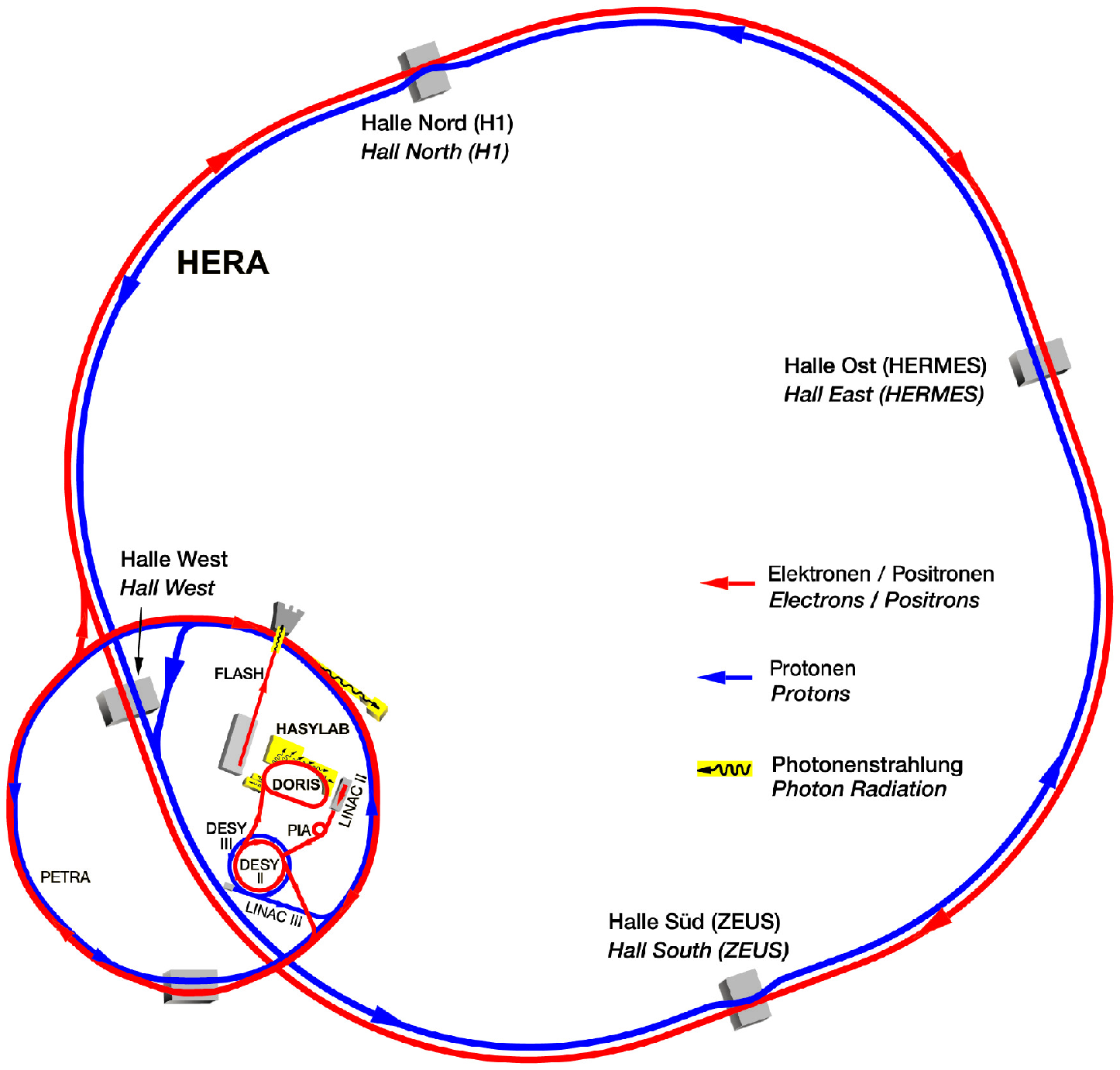}
 \adjincludegraphics[width=0.5\linewidth]
 {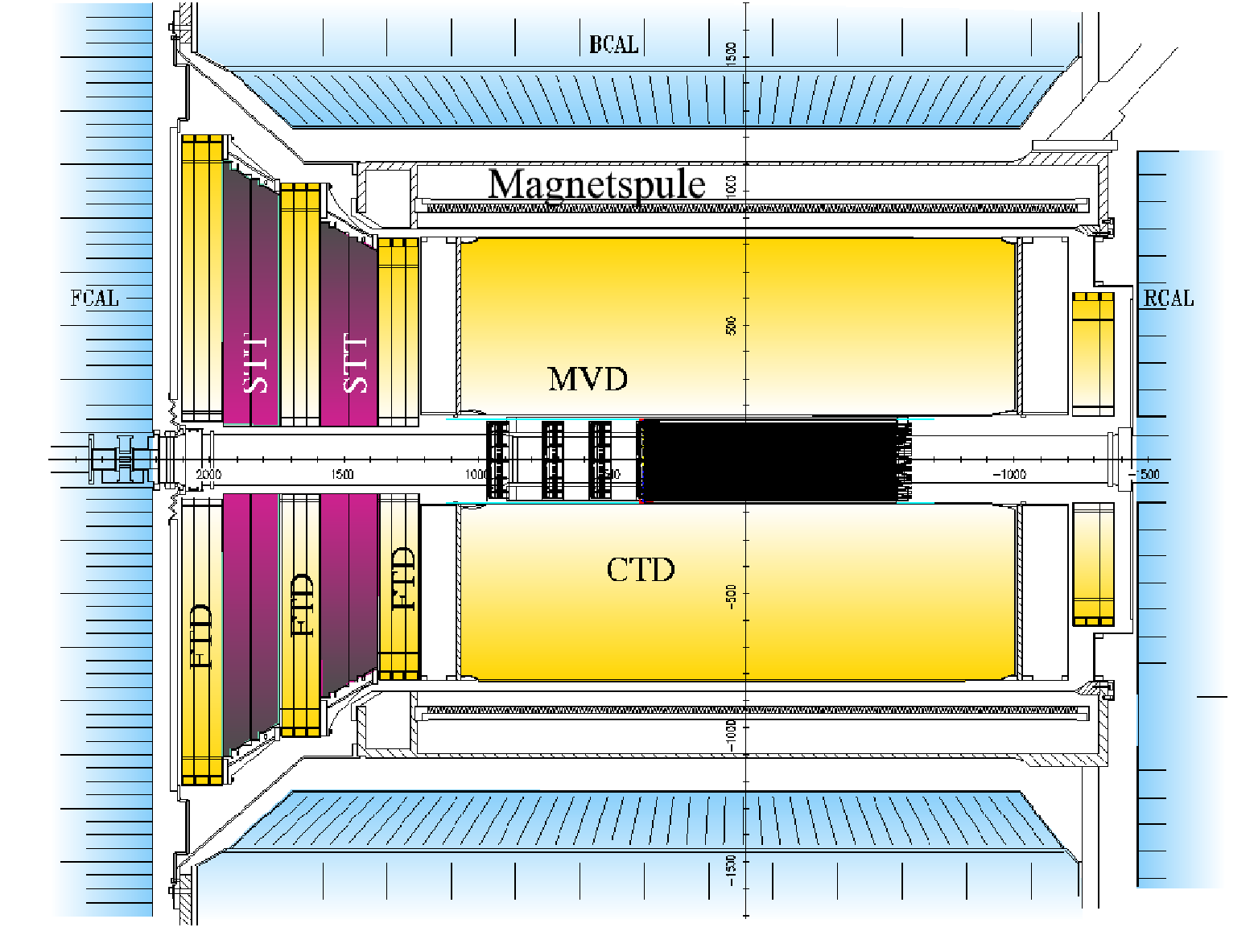}
 \end{center}
 \caption[HERA and ZEUS]{
 Schematic view of \HERA accelerating ring, preacceleration facilities 
 and experimental holes (left) and general schematic  view of \ZEUS 
 tracking system (right)~\protect\cite{Schonberg:2010zza}.}
 \label{zeusdf}
 \end{figure}

 \ZEUS was an experiment taking data in $ep$   collisions at \HERA.
 A full description of the original \ZEUS detector design can be found 
 elsewhere~\cite{zeus:1993:bluebook}. Briefly,
 \ZEUS was a  general purpose detector developed for a broad range of 
 physics studies. It had a size of $12\,{\rm m} \times 
11\,{\rm m} \times 20\,{\rm m}$ and a weight of 3600$\,\rm{tons}$. The 
 major  components of \ZEUS included a system for charged particle 
 tracking for polar angles 7$^\circ$ $<\theta<$ 170$^\circ$ ~ within a 
 high-field superconducting solenoid, a high resolution depleted-uranium
 calorimeter and a  tracking system for muons. The detector was completed
 by dedicated forward detectors and a luminosity monitor.

 In total \ZEUS took more than $360$ billion events of $ep$ collisions
 which corresponds to a total luminosity close to $0.5\fbi$.
 \section{Physics case }
\begin{figure}[!htb]
 \floatbox[{\capbeside\thisfloatsetup{capbesideposition={right,top},
 capbesidewidth=0.55\linewidth}}]{figure}[\FBwidth]
 {\caption[Number of ZEUS publications per year]{The number of \ZEUS 
 published and anticipated publications per year. The colours and grey 
 shades indicate the usage of data from the different data taking 
 periods, and the contribution from results produced from the  preserved data. 
 The corresponding numbers from the four LEP experiments combined, shifted
 forward by 7 years to account for the difference in end of data taking 
 and normalised to the period 1997-2009 for \ZEUS. The excess of the \ZEUS 
 distribution over the normalised LEP distribution beyond the end of 
 funding in 2014 can be considered as a result of dedicated data 
 preservation. 
 \label{FZEUSPUB} }}
 {\adjincludegraphics[width=0.85\linewidth,height=1.1\linewidth,trim={ {0.1\Height} {0.05\Width} 
 {0.17\Height}   {0.1\Width}}, clip=true]
 {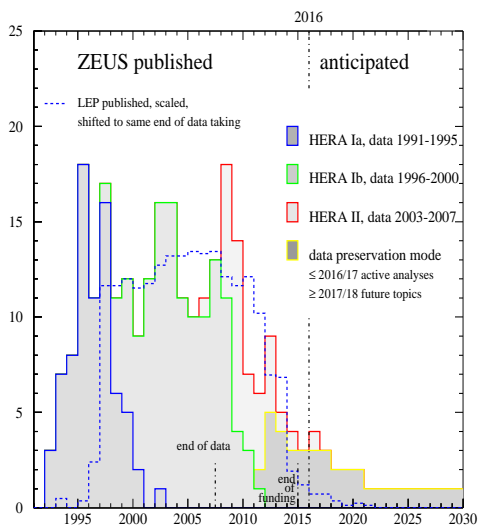}}
 \end{figure}

 After the end of funding period in 2014, the \ZEUS collaboration has 
 produced numerous publications using the preserved data 
 (see Fig.~\ref{FZEUSPUB}). As of 2016 the collaboration actively 
 produces valuable physics 
 results~\cite{Abramowicz:2016xln,Abramowicz:2016ztw,Abramowicz:2016xzf,Abramowicz:2016ext}. 
 It is foreseen that many valuable results can be  obtained  with the 
 preserved \ZEUS data~\cite{Bacchetta:2016rdn,Geiser:2015pwp}.
 The list of analysis topics is discussed in detail in
 Refs.~\cite{Bacchetta:2016rdn,Geiser:2015pwp}. Briefly, some topics of 
 interest are: the proton structure (e.g. $F_2$ and $F_L$ and the 
 strangeness content of the proton), combination of various diffraction 
 measurements, studies of jets and event shapes with N(N)LO predictions,
 studies of photon structure,  prompt photons, measurements of 
 electroweak couplings, searches of instantons.
 \section{Common Ntuples}
 Before the end of 2012, the \ZEUS analysis model was using the data stored in 
 Mini Data Summary Tapes (MDST) in ADAMO format. To perform an  
 analysis a large set of \ZEUS-specific software together with its 
 customisation was needed. It was not intended to be maintained when the 
 available manpower was reduced. Instead, the \ZEUS analysis software was 
 used to create common-usage ntuples for real and Monte Carlo(\MC) data. These are 
 referred to as Common Ntuples(CN) and, by their  structure,  are  simple  
 ROOT~\cite{Antcheva:2011zz}\footnote{PAW files with an identical 
 information were created as well}  files. 
 They contain the necessary information to perform ongoing and 
 planned future analyses; all ongoing and most of the recently published 
 analyses actively use them. 

 \CN are backward compatible as these are based on the existing 
 documented ntuple scheme. This makes use of the expertise and knowledge 
  accumulated in \ZEUS analyses over the last decades.  As ROOT is the 
  main analysis tool for LHC it is expected to survive in a backwards 
  compatible way for a very long time. 
 \section{Bit preservation}
 \label{sec:bit}
 One of the most important  tasks of the \ZEUS data 
 preservation is to save the data bits. The \ZEUS data selected for the 
 preservation consist of the \CN (ROOT/some PAW) and 
 corresponding log files, in total $\approx250{\rm Tb}$ in $1.1$ million files. The data is 
 stored in DESY\footnote{http://www.desy.de, Deutsches Elektronen 
 Synchrotron, Notkestr. 85, Hamburg DE-22607}  and 
 MPCDF\footnote{http://www.mpcdf.mpg.de, Max-Planck Computing 
 and Data Facility, Gie{\ss}enbachstr. 2, Garching DE-85748}.

 In the DESY IT centre the data is stored in the DESY DPHEP 
 dCache~\cite{dccp} system. The system has two instances with identical 
 content: the disk pool system and backup archive tape system. 
 The disk pools can be accessed with a very low latency from the DESY 
 NAF\footnote{http://naf.desy.de/, National Analysis Facility} computing 
 facility and two dedicated machines inside DESY. 
 A DESY/\ZEUS account is required for it. 

 In MPCDF centre the data is stored in the tape system and in the disk 
 pool. The disk pool system contains only a fraction of the data stored 
 in the  archive tape system, namely the data and most important Monte 
 Carlo simulated samples. On a request any specific part of the archived 
 data can be transferred to the disk pool. The disk pools can be 
 accessed with a very low latency from the 
 MPCDF/MPP\footnote{http://www.mpp.mpg.de, Max-Planck Institut f\"{u}r Physik, 
 F\"{o}hringer Ring 6, M\"{u}nchen DE-80805}  machines or with multiple 
 protocols worldwide. For the former a MPCDF/MPP  account with appropriate 
 permissions is needed. For the access to the data worldwide a grid 
 certificate with the \ZEUS Grid Virtual Organisation~\cite{Stadie:2006nq} membership is 
 needed. 
 \section{Software preservation}
 \label{sec:analysis}
 For an analysis of \ZEUS data an appropriate software is needed.
 In this section  the  most important software packages for the \ZEUS 
 analysis with the preserved data are described: the analysis framework, 
 event display, file catalogue and \MC production packages.

 An {\bf analysis framework} is  a core analysis software.
 It is made to read, analyse the data and perform the calculations.
 Before switching to the data preservation mode of analysis,
 considerable efforts were put by the \ZEUS collaboration to minimise 
 the amount of software required for the analysis framework.
 As a result, the standard ROOT package 
 is the framework for the \ZEUS analysis. It is documented in detail
 elsewhere~\cite{Antcheva:2011zz}.

\begin{figure}[htbp]
 \centering
 \includegraphics[width=\linewidth,height=0.36\linewidth]
 {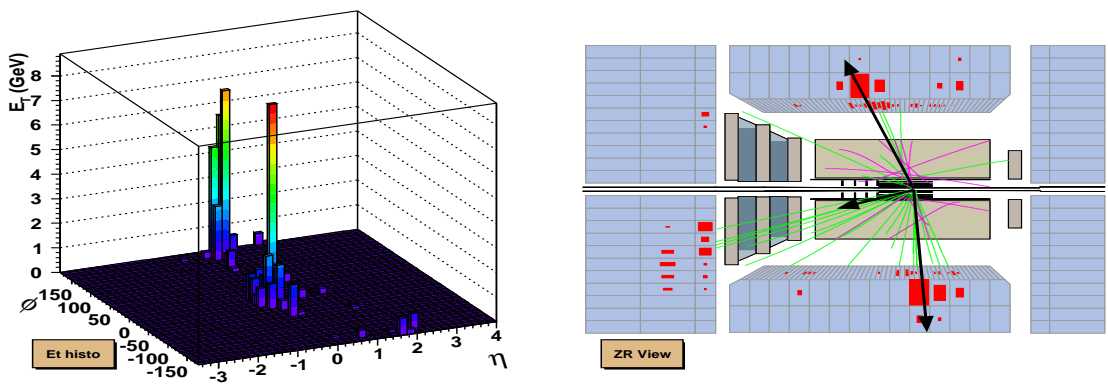}
 \caption[One of selected events from $Z^0$-boson analysis]
 {One of the selected events from the $Z^0$-boson 
 analysis~\cite{Abramowicz:2012qy}. The transverse energy, 
 $E_T(\eta,\phi)$, distribution in the CAL sells (left) and ZR event 
 view (right) are shown.}
 \label{fig:zevis_z0bosonevent}
 \end{figure}

 An {\bf event display } is a program for the visualisation of particle 
 collision event data in detector. It serves to 
 adjust and check complicated reconstruction techniques by looking 
 at single typical events and to  produce illustrations of sophisticated 
 analysis techniques for talks and papers (see Fig.~\ref{fig:zevis_z0bosonevent}). 
 During the \HERAII period  an event display called \ZEVIS has been 
 developed using ROOT libraries. Originally it was a 
 client-server application where  the dedicated server performed  the 
 reconstruction from MDSTs creating a ROOT file with the required information.
 The file was transmitted to the client  via the network for the visualisation. 
 In the context of the 
 data preservation the client  was modified to read the 
 information from the \CN.
  The detailed description of the package 
 is given in  Ref.~\cite{Zenaiev:2015qea}. 
 
 A {\bf filecatalog} is a database  with an information on the available 
 data samples.  The   \ZEUS filecatalog,  called CNINFO, provides an 
 information on the available data and Monte 
 Carlo samples in \CN format: the number of events, the data 
 taking period, \CN version, run numbers etc. The information 
 can be  extracted from the database with corresponding utilities. The 
 database itself is stored in SQLite3~\cite{sqlite} and plain text formats. An option 
 to edit (e.g.\ add new \MC samples) is foreseen.

 Some of the future analyses might require the generation  of new \MC samples.
 The {\bf \MC generators} used during active data taking are deeply 
 integrated into \ZEUS software and produced the events in the 
 \ZEUS-specific event record format. 
 As of 2016 it is possible to use these for the \MC events generation,
 but an integration of a new  
 generator to this scheme requires significant efforts or is not
 possible at all. To solve this problem, an utility for a 
 conversion of the output of \MC generators  to \ZEUS-specific 
 format was developed. The conversion is possible for multiple formats, 
 including HEPMC2~\cite{Dobbs:2001ck}, HEPMC3~\cite{hepmc3} and 
 HEPEVT~\cite{Sjostrand:1989}. In the data preservation mode of analysis 
 this is the favourable 
 way to produce \MC events and the only way to do it with the new \MC
 generators.

 For the purpose of \ZEUS DP  the full  software 
 chain for the detector simulation and reconstruction of \MC simulated 
 events, completely independent from the original analysis environment
  has been prepared.  The software chain is called \ZEUS MC Standalone 
  Package ({\bf ZMCSP}) and is based on the software used for the \ZEUS \ \MC simulations on 
 the Grid~\cite{Stadie:2006nq}.
 Briefly, the package is a tarball that contains all the 
 libraries and executable needed for the production of the \MC simulated 
 samples. The input for the package is the \MC generated events  
 and the output is the 
 \CN files (ROOT and/or PAW) described above. 
 The software chain has been designed to be as much as possible 
 independent from the host system. This makes it possible to use ZMCSP outside the \DESY
 environment
 and benefit from  other computing resources available   
 in member institutes and elsewhere worldwide. 
 Hereby, for the \MC event generation, the  Grid technology is especially 
 attractive as it has been successfully exploited 
 for many years for  the same tasks in  \ZEUS~\cite{Stadie:2006nq}. 

 As of 2016, the described software for  \ZEUS analysis 
 can be used on many contemporary systems.
 The discussed long-term solutions for a system that can run the software
 are the NAF facility and virtual machines. 
 Briefly, the first option is to use the standard analysis farm (or a part of it) 
 that will be maintained at DESY with an option of access to CN on a long term.
 The second option is 
 a customised CentOS~\cite{centos} installation image suitable for various  
 (virtual) machines with the relevant software. The image and the 
 relevant instructions are provided on request (see Sec.~\ref{sec:accessrules}).  
 \section{Documentation preservation}
 The main efforts to preserve the digital and paper documentation of the 
 \ZEUS experiment were
 conducted in DESY. A brief description of the activities can be found 
 in Ref.~\cite{Malka:2012hg}.
 Concepts  for  long-term  preservation  of  the  \ZEUS  digital  
 documentation  were 
 worked out in collaboration with the \DESY IT division and DESY library. 
 These are  based on the 
 \INSPIRE\footnote{http://inspirehep.net}  system  and  centrally 
 maintained  web  servers\footnote{http://www-zeus.desy.de, http://wwwzeus.mpp.mpg.de}. 
 As of 2016 all of the ZEUS internal notes  are available (with password) 
 on \INSPIRE.
 The archive  of  non-digital  information is hosted  by  the  \DESY  
 library. It includes  all  \ZEUS  notes 
 written  before  1995,  transparencies  presented  at  meetings  
 before  2000,  technical 
 drawings etc. It is also foreseen that 
 in some cases it might be useful to 
 add new or update the existing 
 documentation. 
 \section{Access rules}
 \label{sec:accessrules}
 In the context of the data preservation it is important to clarify the 
 rules for the data and software access.
 The access rules for the \ZEUS data as of 2016 in the data preservation 
 mode are similar to
 these during the active phase of the collaboration.
 In addition, interested analysers from institutes not in \ZEUS can get
 access to the data under conditions to be negotiated with the 
 collaboration spokesperson.
 \section{Conclusions}
 \label{sec:con}
 The \ZEUS data preservation project  has multiple 
 levels and
 many participants. The presented work describes its
 implementation and potential benefits. The ultimate goal is to
 encourage potential collaborators to perform new analysis 
 and continue to deliver valuable physics results.


{\bibliographystyle{ZDP}{\raggedright\bibliography{DIS2016_Andrii_Verbytskyi_proceedings.bib}}}
\vfill\eject  

\end{document}